\documentclass[12pt] {article}

\begin{document}

\input amssym.tex

\title{Remarks on the spherical waves of the Dirac field on de Sitter spacetime}

\author{Ion I. Cot\u aescu,  Radu Racoceanu and
Cosmin Crucean\\ {\it  West University of Timi\c soara,}\\{\it V.
Parvan Ave. 4, RO-300223 Timi\c soara}}

\maketitle

\begin{abstract}
The Shishkin's  solutions of the Dirac equation in spherical moving
frames of the de Sitter spacetime are investigated pointing out the
set of commuting operators whose eigenvalues determine the
integration constants. It is shown that these depend on the usual
angular quantum numbers and, in addition, on the value of the scalar
momentum. With these elements a new result is obtained finding the
system of solutions normalized (in generalized sense) in the scale
of scalar momentum.

Pacs: 04.62.+v
\end{abstract}

\newpage

The Dirac equation on de Sitter spacetime $M$ ( whose cosmological
constant $\Lambda _c$ gives the radius $R=1/\omega=
\sqrt{3/\Lambda_c}$,) has been studied in moving or static local
charts (i.e., natural frames) suitable for separation of variables
leading to significant analytical solutions \cite{OT}-\cite{KT}.

The first spinor solutions on $M$ were obtained in a {\em static}
central chart with spherical coordinates, $\{t_s,r_s,\theta,\phi\}$,
using the diagonal tetrad gauge \cite{OT}. Another gauge of
Cartesian type leads to a system of particular solutions of the
Dirac equation written in terms of the well-known spherical spinors
of special relativity \cite{C1}. All these particular solutions are
eigenspinors of the Hamiltonian operator $H=i\partial_{t_s}$ which
has a continuous energy spectrum. The form of these solutions is too
complicated such that the normalization in the energy scale can not
be done.

Of a particular interest is the {\em moving} frame with the proper
time and spherical space coordinates, $\{t,r,\theta,\phi\}$,
associated to the Cartesian chart $\{t, \vec{x}\}$ with the line
element \cite{BD}
\begin{equation}\label{mssu}
ds^{2}=dt^2 - e^{2\omega t} d\vec{x}^2\,.
\end{equation}
A set of particular spherically symmetric solutions of the Dirac
field in the chart $\{t,r,\theta,\phi\}$ was found by Shishkin
\cite{SHI} using a Cartesian tetrad gauge and a suitable method of
separating variables \cite{ALG}. We note that in the moving charts
the operator $i\partial_{t}$ is no longer a Killing vector field
and, therefore, the separation of variables leads to solutions which
are no energy eigenspinors, their integration constants depending on
other physical quantities. However, to our knowledge, a complete
system of normalized spherical wave solutions of the Dirac equation
was not constructed so far.

The plane wave solutions in the diagonal gauge of the chart $\{t,
\vec{x}\}$ were obtained in \cite{BADU,FGV} but these solutions are
normalized only in the asymptotic approximation. The first complete
system of normalized plane wave solutions in this chart was derived
using a suitable complete set of commuting operators representing
conserved observables associated with the specific isometries of the
de Sitter manifolds \cite{C2}. In this manner all the integration
constants were determined as eigenvalues of the operators of this
set, finding the form of the particular solutions that can be
normalized in generalized sense in the momentum scale.

In this letter we would like to apply the same method for the
spherical waves in the chart $\{t,r,\theta,\phi\}$ trying to
construct normalized solutions as linear combinations of the
Shishkin's ones. Our approach is based on the theory of external
symmetry \cite{ES} which explains the relations among the geometric
symmetries and the operators commuting with the Dirac one,
constructed with the help of the Killing vectors some time ago
\cite{CML}. In fact, these operators are nothing other than the
generators of the spinor representation of the universal covering
group of the isometry group \cite{ES} and, therefore, they
constitute the main physical observables among which we can choose
different sets of commuting operators defining quantum modes. This
method is efficient especially in the case of the de Sitter
spacetime where the high symmetry given by the $SO(4,1)$ isometry
group \cite{SW,WALD} offers one the opportunity of a rich algebra of
conserved operators able to receive a physical meaning.

Given an arbitrary  chart $\{x\}$ of $M$ with coordinates $x^{\mu}$
($\mu, \nu,...= 0,1,2,3$), we have to choose the tetrad fields,
$e_{\hat\mu}(x)$ and $\hat e^{\hat\mu}(x)$ which define the local
(unholonomic) frames and the corresponding coframes. These are
labeled by the local indices, $\hat\mu, \hat\nu,...=0,1,2,3$, and
have the orthonormalization properties \cite{C2} with respect to the
flat metric $\eta=$diag$(1,-1,-1,-1)$. The metric tensor of $M$,
$g_{\mu\nu}=\eta_{\hat\alpha \hat\beta}\hat e^{\hat\alpha}_{\mu}\hat
e^{\hat\beta}_{\nu}$, raises or lowers the Greek indices while for
the local indices (with hat) we have to use the flat metric. In what
follows we consider the Cartesian moving charts  $\{t, \vec{x}\}$,
and $\{t_c, \vec{x}\}$ as well as the corresponding spherical ones,
$\{t,r,\theta,\phi\}$ and $\{t_c,r,\theta,\phi\}$. Here $t_c$ is the
conformal time defined by
\begin{equation}
 \omega t_c =-\, e^{-\omega t}\,,
\end{equation}
which gives the simpler line element  \cite{BD},
\begin{equation}\label{mconf}
ds^{2}=\frac{1}{\omega^2 {t_c}^2}\left({dt_c}^{2}-d\vec{x}^2\right)\,.
\end{equation}
We note that the coordinates of the moving frames are related to
those of the static ones through $e^{-\omega t}=e^{-\omega t_s}\cosh
\omega r_s$ and $r=e^{-\omega t_s}\sinh \omega r_s$ while the
angular variables remain the same.

The theory of the Dirac field $\psi$ minimally coupled with
gravitation can be written simply in the Cartesian gauge where  the
non-vanishing tetrad components  are  \cite{SHI}
\begin{equation}\label{tt}
e^{0}_{0}=-\omega t_{c}\,, \quad e^{i}_{j}=-\delta^{i}_{j}\,\omega t_c
\,,\quad
\hat e^{0}_{0}=-\frac{1}{\omega t_{c}}\,, \quad \hat e^{i}_{j}=-\delta^{i}_{j}\,
\frac{1}{\omega t_c}\,.
\end{equation}
In this gauge the Dirac operator reads \cite{C2}
\begin{eqnarray}\label{ED1}
E_D&=&-i\omega t_c\left(\gamma^0\partial_{t_{c}}+\gamma^i\partial_i\right)
+\frac{3i\omega}{2}\gamma^{0} \nonumber\\
&=&i\gamma^0\partial_{t}+ie^{-\omega t}\gamma^i\partial_i
+\frac{3i\omega}{2}\gamma^{0}\,,
\end{eqnarray}
where the $\gamma$ matrices satisfy $ \{
\gamma^{\hat\alpha},\gamma^{\hat\beta}
\}=2\eta^{\hat\alpha\hat\beta}$ and $
S^{\hat\mu\hat\nu}=\frac{i}{4}[\gamma^{\hat\mu},\gamma^{\hat\nu}]$.
The conserved operators commuting with $E_D$ are the generators of
the external symmetry group $S(M)$ which is just the universal
covering group of the isometry group $I(M)=SO(4,1)$ \cite{ES}. In
Ref. \cite{C2} we pointed out that the Hamiltonian operator $H$, the
components of the {\em momentum}, $P^i$, and those of the total
angular momentum,  $J^i=\varepsilon_{ijk}J_{jk}/2$, are the
following basis-generators of $S(M)$
\begin{eqnarray}
P^{i}&=&-i\partial_{i}\label{Gip}\\
H&=&-i\omega(t_{c}\partial_{t_{c}}+x^{i}\partial_i)
\label{Gi}\\
J_{ij}&=&-i(x^i\partial_j-x^j\partial_i)+S_{ij}\,,
\end{eqnarray}
remaining with three more basis-generators which do not have an
immediate physical significance \cite{C2}. In the chart,
$\{t,\vec{x}\}$, the operators $\vec{P}$ and
$\vec{J}=\vec{L}+\vec{S}$ (with $\vec{L}=\vec{x}\times \vec{P}$)
keep their forms while the Hamiltonian operator becomes
$H=i\partial_{t}+\omega\, \vec{x}\cdot\vec{P}$. We observe that the
presence of the external gravitational field ($\omega\not = 0$)
leads to the commutation rules
\begin{equation}\label{cHP}
[H,\,P^i]=i\omega P^i \,,
\end{equation}
which prevent one to diagonalize simultaneously the operators $H$
and $P^i$. For this reason,  the plane wave solutions of \cite{C2}
were derived as common eigenspinors of the complete set of commuting
operators $\{E_D,P^i, W=\vec{P}\cdot\vec{S}\}$.

In these circumstances it is natural to consider the spherical modes
defined by the common eigenspinors of the complete set
$\{E_D,\vec{P}^2, \vec{J}^2,K, J_3\}$ where $\vec{P}^2$ plays the
role of $H$ in static frames \cite{C1}. We remind the reader that
the operator $K=\gamma^0(2\vec{L}\cdot\vec{S}+1)$ concentrates the
action of all the angular operators. Our purpose is to write down
the particular solutions of the Dirac equation $ E_D \psi = m\psi$,
of mass $m$, in such a way that $\psi= \psi _{p,\kappa_j, m_j}$  be
a common eigenspinor of the above set of commuting operators,
corresponding to the eigenvalues $\{m,p^2, j(j+1), -\kappa_j, m_j\}$
where $\kappa_j=\pm (j+\frac{1}{2})$  \cite{TH} while $p$ is the
value of the {\em scalar momentum}. In addition, we require these
solutions to be normalized with respect to the time-independent
relativistic scalar product defined as \cite{C2}
\begin{equation}\label{ps}
\left<\psi, \psi'\right>=\int d^3 x\, e^{3\omega
t}\,\overline{\psi}(t,\vec{x})\gamma^0\psi'(t,\vec{x})\,,
\end{equation}
in the chart $\{t,\vec{x}\}$.

For solving the above eigenvalue problems it is convenient to start
with the chart $\{x\}=\{t_c,r,\theta,\phi\}$ looking for particular
solutions of the form
\begin{equation}
\psi_{p,\kappa_j, m_j}(x)=\frac{(-\omega
t_c)^{\frac{3}{2}}}{r}\left[f^+_{p,\kappa_j}(t_c,r)
\Phi^+_{m_j,\kappa_j}(\theta,\phi)
+f^-_{p,\kappa_j}(t_c,r)\Phi^-_{m_j,\kappa_j}(\theta,\phi)\right]
\end{equation}
where $\Phi^{\pm}_{m_j,\kappa_j}$ are the usual normalized spherical
spinors of special relativity that solve the eigenvalue problems of
the operators $\vec{J}^2$, $K$ and $J_3$ \cite{TH}. Then, denoting
by $k=m/\omega$, after a little calculation, we arrive ar the pair
of equations
\begin{equation}\label{rad}
\left(\pm i\partial_{t_c}+\frac{k}{t_c}
\right)f^{\pm}_{p,\kappa_j}(t_c,r)= \left( -\partial_r\pm
\frac{\kappa_j}{r}\right)f^{\mp}_{p,\kappa_j}(t_c,r)\,,
\end{equation}
resulted from the Dirac one. In addition,  the eigenvalue problem of
$\vec{P}^2$ leads to the supplemental radial equations
\begin{equation}\label{PP}
\left[-\partial_r^2+\frac{\kappa_j(\kappa_j\pm
1)}{r^2}\right]f^{\pm}_{p,\kappa_j}(t_c,r)=p^2
f^{\pm}_{p,\kappa_j}(t_c,r)\,,
\end{equation}
since the spinors $\Phi^{\pm}_{m_j,\kappa_j}$ are eigenfunctions of
$\vec{L}^2$ corresponding to the eigenvalues $\kappa_j(\kappa_j\pm
1)$. Eqs. (\ref{rad}) and (\ref{PP}) can be solved separating the
variables,
\begin{equation}
f^{\pm}_{p,\kappa_j}(t_c,r)=\tau^{\pm}_p(t_c)
\rho_{p,\kappa_j}^{\pm}(r)\,,
\end{equation}
and finding  that the new functions must satisfy
\begin{eqnarray}
\left(\pm i\partial_{t_c}+\frac{k}{t_c}
\right)\tau^{\pm}_p(t_c)&=&\pm
p \,\tau^{\mp}_p(t_c)\,,\\
\left(\pm
\partial_{r}+\frac{\kappa_j}{r} \right)\rho^{\pm}_{p,\kappa_j}(r)&=& p
\,\rho^{\mp}_{p,\kappa_j}(r)\,.
\end{eqnarray}
These equations have to be solved in terms of Bessel functions
\cite{AS} as in \cite{SHI}. The advantage of our method is to point
out that there is only one additional integration constant, $p$,
which is a continuous parameter with a precise physical meaning
(i.e., the scalar momentum).

However, our main problem is to find the normalized solutions with
respect to the scalar product (\ref{ps}). We specify that these
solutions are not square integrable since the spectrum of
$\vec{P}^2$ is continuous. Therefore, we must look for a system of
spinors $\psi_{p,\kappa_j, m_j}$ normalized in the generalized sense
in the scale of the scalar momentum $p$. First, we choose the radial
functions as in \cite{SHI},
\begin{equation}
\rho^{\pm}_{p,\kappa_j}(r)=\sqrt{pr}\,J_{|\kappa_j\pm
\frac{1}{2}|}(pr)\,.
\end{equation}
Furthermore, we observe that the functions $\tau^{\pm}$ must produce
a similar time modulation as in the case of the normalized plan wave
solutions of  \cite{C2}. Consequently, we denote
$\nu_{\pm}=\frac{1}{2}\pm ik$ and write the functions $\tau^{\pm}$
in terms of Hankel functions \cite{AS} as
\begin{equation}
\tau^{\pm}_p(t_c)= N\sqrt{-p t_{c}}\,e^{\pm\pi
k/2}H^{(1)}_{\nu_{\mp}}(-p t_{c})
\end{equation}
where $N$ is a normalization factor. Finally, summarizing these
results in the chart $\{x\}=\{t,r,\theta,\phi\}$ and matching the
factor $N$ we find the definitive form of the normalized spinors
\begin{eqnarray}\label{sol}
&&\psi_{p,\kappa_j, m_j}(x)=\frac{p}{2} \sqrt{\frac{\pi}{\omega
r}}\,e^{-2\omega t}\left[e^{\pi
k/2}H^{(1)}_{\nu_-}(\textstyle{\frac{p}{\omega}} e^{-\omega t})
J_{|\kappa_j+\frac{1}{2}|}(pr)\Phi^+_{m_j,\kappa_j}(\theta,\phi)\right.\nonumber \\
&&~~~~~~~~~~~~~~~~~~~~~~~~~\left.+e^{-\pi
k/2}H^{(1)}_{\nu_+}(\textstyle{\frac{p}{\omega}} e^{-\omega t})
J_{|\kappa_j-\frac{1}{2}|}(pr)\Phi^-_{m_j,\kappa_j}(\theta,\phi)
\right]
\end{eqnarray}
that satisfy the Dirac equation and are common eigenspinors of the
operators $\vec{P}^2, \vec{J}^2, K$ and $J_3$. Taking into account
that \cite{LL}
\begin{equation}
\int_{0}^{\infty}\rho_{p,\kappa_j}(pr)
\rho_{p',\kappa_j}(p'r)dr=\delta(p-p')\,,
\end{equation}
and using the properties of Hankel functions mentioned in \cite{C2}
we obtain the orthonormalization rule
\begin{equation}
\left<\psi_{p,\kappa_j, m_j}, \psi_{p',\kappa'_{j'},
m'_j}\right>=\delta(p-p')\delta_{j,j'}\delta_{\kappa_j,\kappa'_{j}}\delta_{m_j,m'_j}\,.
\end{equation}

According to the conventions of \cite{C2} we can say that the
particular solutions (\ref{sol}) are of positive frequencies and,
therefore, these describe the quantum modes of the Dirac particles.
The spinors of negative frequencies, corresponding to antiparticles,
will be obtained using the charge conjugation as in \cite{C2}. In
this way one may obtain a complete system of orthonormalized spinors
which could be the starting point to the canonical quantization of
the Dirac field in spherical moving frames.


\end{document}